\documentclass[12pt]{article}
\usepackage{graphicx}
\usepackage{cite}

\def \bea{\begin{eqnarray}}
\def \beq{\begin{equation}}
\def \b{{\cal B}}
\def \bo{B^0}

\def \ckp{C_{K \pi}}
\def \cpp{C_{\pi^+ \pi^-}}
\def \eea{\end{eqnarray}}
\def \eeq{\end{equation}}

\def \ob{\overline{B}^0}
\def \s{\sqrt{2}}
\def \skp{S_{K \pi}}
\def \spp{S_{\pi^+ \pi^-}}

\def \3half{\frac{3}{2}}
\def \ko{K^0}
\def \ok{\overline{K}^0}

\begin{document}
\begin{flushright}
EFI 08-21 \\
July 2008 \\
arXiv:0807.3080 \\
\end{flushright}
\centerline{\bf Implications for CP asymmetries of improved data 
on $B \to K^0 \pi^0$}
\bigskip
\centerline{Michael Gronau\footnote{On sabbatical leave from the Physics 
Department, Technion, Haifa 32000, Israel.} and Jonathan L. Rosner}
\medskip
\centerline{\it Enrico Fermi Institute and Department of Physics,
 University of Chicago} 
\centerline{\it Chicago, IL 60637, U.S.A.} 
\bigskip
\begin{quote}
The decay $B^0 \to K^0 \pi^0$, dominated by a $b \to s$ penguin amplitude,
holds the potential for exhibiting new physics in this amplitude.  In the pure
QCD penguin limit one expects $\ckp = 0$ and $\skp = \sin 2 \beta$ for the
coefficients of $\cos \Delta m t$ and $\sin \Delta m t$ in the time-dependent
CP asymmetry.  Small non-penguin contributions lead to corrections to these
expressions which are calculated in terms of isospin-related $B\to K\pi$ rates
and asymmetries, using information about strong phases from experiment.  We
study the prospects for incisive tests of the Standard Model through
examination of these corrections.
We update a prediction $\ckp=0.15\pm 0.04$,
pointing out the sensitivity of a prediction
$\skp\approx 1$ to the measured branching ratio for $B^0\to K^0\pi^0$ and to
other observables.
\end{quote}

\leftline{\qquad PACS codes:  12.15.Hh, 12.15.Ji, 13.25.Hw, 14.40.Nd} 

\medskip
One of the most challenging CP asymmetry measurements in $B$ meson decays has
involved the coefficients $\ckp$ and $\skp$ in the time-dependent 
asymmetry measured in $B^0\to K_S\pi^0$~\cite{Gronau:1989ia}
\beq
A(t) = \frac{\Gamma(\ob(t) \to \ok \pi^0) - \Gamma(\bo(t) \to \ko \pi^0)}
      {\Gamma(\ob(t) \to \ok \pi^0) + \Gamma(\bo(t) \to \ko \pi^0)}
     = - \ckp \cos(\Delta m t) + \skp \sin(\Delta m t)~.
\eeq
The parameter $\ckp$ is related to the direct CP asymmetry by 
$\ckp\equiv -A_{CP}(B^0\to K^0\pi^0)$.
The decay $B^0 \to K^0 \pi^0$ is expected to be dominated by the $b \to s$
penguin amplitude and thus is a good place to look for any new physics that may
arise in this amplitude \cite{Gronau:1996rv,Grossman:1996ke,London:1997zk}. 
In the pure QCD penguin limit one expects $\ckp = 0$ and $\skp = \sin 2 \beta$,
respectively, where $\beta = (21.5 \pm 1.0)^\circ$~\cite{HFAG} is an angle in
the unitarity triangle.  Accounting for small non-penguin contributions leads
to corrections to these expressions, which are calculable in terms of
isospin-related $B\to K\pi$ decay rates and asymmetries.  
In this Letter we study the prospects for incisive tests of the Standard Model 
through examination of these corrections.
We update a prediction $\ckp=0.15\pm 0.04$ and 
point out the sensitivity of a recent theoretical prediction  
$\skp\approx 1$~\cite{Fleischer:2008wb} to the branching ratio for 
$B^0\to K^0\pi^0$ and to other observables.

\begin{table}
\caption{Measurements of $\ckp$ and $\skp$.
\label{tab:ckpskp}}
\begin{center}
\begin{tabular}{c c c} \hline \hline
Ref.\ & $\ckp$ & $\skp$ \\ \hline
BaBar \cite{Aubert:2007mgb} & $0.24 \pm 0.15 \pm 0.03$ & $0.40 \pm 0.23 \pm 0.03$ \\
Belle \cite{Abe:2006gy} & $0.05 \pm 0.14 \pm 0.05$ & $0.33 \pm 0.35 \pm 0.08$ \\
Average \cite{HFAG} & $0.14 \pm 0.11$ & $0.38 \pm 0.19$ \\ \hline
\end{tabular}
\end{center}
\end{table}
%
\begin{table}
\caption{CP-averaged branching ratios and CP rate asymmetries for $B \to K \pi$
decays and $B^+\to\pi^+\pi^0$, based on averages in Ref.\ \cite{HFAG}.
\label{tab:data}}
\begin{center}
\begin{tabular}{c c c} \hline \hline
       Mode         &    Branching      &     $A_{CP}$    \\
                    & ratio ($10^{-6}$) &                 \\ \hline
$B^0 \to K^+ \pi^-$ &  $19.4 \pm 0.6$   & $-0.097\pm0.012$ \\
$B^0 \to K^0 \pi^0$ &   $9.8 \pm 0.6$   & $-0.14\pm 0.11$ \\
$B^+ \to K^0 \pi^+$ &  $23.1 \pm 1.0$   & $0.009\pm0.025$ \\
$B^+ \to K^+ \pi^0$ &  $12.9 \pm 0.6$   & $0.050\pm0.025$ \\
$B^+\to \pi^+\pi^0$ & $5.59^{+0.41}_{-0.40}$ & $0.06\pm 0.05$\\
\hline \hline
\end{tabular}
\end{center}
\end{table}
The current status of measurements of $\ckp$ and $\skp$ is summarized in
Table \ref{tab:ckpskp}.  The value of $\ckp$ is consistent with the
pure-penguin value of zero, while that of $\skp$ is $1.6 \sigma$ below the
pure-penguin value of $\sin 2 \beta = 0.681 \pm 0.025$.

A sum rule for direct CP asymmetries in $B \to K \pi$ decays has been derived
purely on the basis of the $\Delta I = 0$ property of the
dominant penguin amplitude, using an isospin quadrangle relation among
the four $B\to K\pi$ decay amplitudes which depend also on two 
$\Delta I=1$ amplitudes~\cite{Nir:1991cu,Gronau:1991dq}:
\beq\label{Quad}
A(B^0\to K^+\pi^-) + \s A(B^0\to K^0\pi^0) = A(B^+\to K^0\pi^+) + \s A(B^+\to K^+\pi^0)~.
\eeq
In its most precise form the sum rule relates the four CP rate differences~\cite{Gronau:2005kz}, 
\beq\label{SRDelta}
\Delta(K^+ \pi^-) + \Delta(K^0 \pi^+) = 2 \Delta(K^+ \pi^0) +
2\Delta(K^0 \pi^0)~,
\eeq
where one defines
\beq
\Delta(f) \equiv \Gamma(\bar B \to \bar f) - \Gamma(B \to f)~.
\eeq
This sum rule includes  interference terms  of the dominant penguin 
amplitude with all small non-penguin contributions. A few very small quadratic terms 
representing interference of tree and electroweak penguin amplitudes vanish in 
the SU(3) and heavy quark limits~~\cite{Gronau:2005kz}.

Using the decay branching ratios and CP asymmetries summarized in Table
\ref{tab:data}~\cite{HFAG} and the known lifetime ratio $\tau(B^+)/
\tau(B^0) = 1.071 \pm 0.009$ \cite{HFAG}, one can use this relation to
solve for the least-well-known quantity $\Delta(K^0 \pi^0)$, implying
\beq\label{pred}
A_{CP}(K^0 \pi^0) = -0.148 \pm 0.044~.
\eeq 
The error on the right-hand-side is dominated by the current experimental
errors in $A_{CP}(K^0 \pi^+)$ and $A_{CP}(K^+ \pi^0)$. The prediction
(\ref{pred}) following from (\ref{SRDelta}) involves a smaller theoretical
uncertainty at a percent level from quadratic terms describing the interference
of small non-penguin amplitudes. Verification of this prediction would provide
evidence that non-penguin amplitudes behave as expected in the Standard Model.
[If one uses the corresponding sum rule for CP asymmetries,
\beq
A_{CP}(K^+ \pi^-) + A_{CP}(K^0 \pi^+) = A_{CP}(K^+ \pi^0) +
A_{CP}(K^0 \pi^0)~,
\eeq
one predicts $A_{CP}(K^0 \pi^0) = - 0.138 \pm 0.037$.  Using this
relation with $A_{CP}(K^0 \pi^+) = 0$, as expected since $B^+ \to K^0
\pi^+$ should be dominated by a penguin amplitude with only a very small
annihilation contribution \cite{Gronau:2005gz}, one predicts $A_{CP}(K^0
\pi^0) = -0.147 \pm 0.028$.]

Non-penguin amplitudes are generally agreed to increase $\skp$ from its
pure-penguin value of $\sin 2 \beta = 0.681 \pm 0.025$ by a modest amount,
generally to 0.8 or below 
\cite{Chiang:2004nm,Cheng:2005bg,Beneke:2005pu,Williamson:2006hb}.  
Model-independent bounds
using flavor SU(3)~\cite{Gronau:2003kx,Gronau:2006qh} also favor 
at most a deviation of 0.2 from the pure-penguin
value.  An exception is noted in the treatments of Refs.\ \cite{Buras:2003dj}
and \cite{Buras:2004ub}, and most recently in Ref.\ \cite{Fleischer:2008wb},
where a relation between $\ckp$ and $\skp$ was studied implying a value 
$\skp=0.99$ for the central value measured for $\ckp$. 
A geometrical construction is
performed which illustrates the way in which such a large value arises.

An aspect of the prediction of $\skp \simeq 0.99$ which has not been
sufficiently stressed is its extreme sensitivity to the branching ratio
$\b(B^0 \to K^0 \pi^0)$.  In the present Letter we analyze the sensitivity
of $\skp$ to this and other observables within the Standard Model, and
highlight those measurements which would shed light on the presence of new
physics.  In order to restrict the range allowed for $\skp$ in the Standard
Model one needs certain information about strong phases. Theoretical
calculations of strong phases in $B\to K\pi$ based on $1/m_b$ expansions are
known to fail, most  likely because of long distance charming penguin 
contributions~\cite{Ciuchini:1997rj,Jain:2007dy}. We propose to obtain the
necessary information about strong phases directly from experiments. 
Somewhat different but not completely independent arguments were presented
in Ref.~\cite{Fleischer:2008wb}.

The $B \to K \pi$ amplitudes may be decomposed into contributions from
various amplitudes as follows~\cite{Gronau:1994rj,Gronau:1995hn}:
\bea
A_{+-} & \equiv & A(B^0 \to K^+ \pi^-) = -(p+t)~~~, \nonumber \\
A_{00} & \equiv & \s A(B^0 \to K^0 \pi^0) = p-c~~~, \nonumber \\
A_{0+} & \equiv & A(B^+ \to K^0 \pi^+) = p + A~~~, \nonumber \\
A_{+0} & \equiv & \s A(B^+ \to K^+ \pi^0) = -(p+t+c+A)~~~,
\label{eqn:decomp}
\eea
\beq
t \equiv T + P_{\rm EW}^C~,~~ c \equiv C + P_{\rm EW}~,~~
p \equiv P - \frac{1}{3} P_{\rm EW}^C~~~.
\eeq
The terms $T, C$ and $A$ represent color-favored and color-suppressed tree
amplitudes and a small annihilation term, while $P$ stands for a gluonic
penguin amplitude.  Color-favored and color-suppressed electroweak penguin
amplitudes are represented by $P_{\rm EW}$ and $P_{\rm EW}^C$.  
The sums of the first two and last two amplitudes in Eq.\ (\ref{eqn:decomp})
are equal [see Eq.~(\ref{Quad})] and both correspond to an amplitude $A_{3/2}$
for a $K \pi$ state with isospin $I_{K \pi} = 3/2$ \cite{Nir:1991cu,Gronau:1991dq}:
$$
A(B^0 \to K^+ \pi^-) + \s A(B^0 \to K^0 \pi^0) = 
A(B^+ \to K^0 \pi^+) + \s A(B^+ \to K^+ \pi^0)
$$
\beq
 =  -(t+c) = -(T + C + P_{\rm EW}^C + P_{\rm EW}) = A_{3/2}~.
\label{eqn:tcsum}
\eeq

The contribution $-(T+C)$ to $A_{3/2}$ has a magnitude which can be obtained
from the decay $B^+ \to \pi^+ \pi^0$ via flavor SU(3) \cite{Gronau:1994bn},
\beq \label{eqn:xitc}
|T+C|=\s\frac{V_{us}}{V_{ud}}\frac{f_K}{f_\pi}\xi_{T+C}|A(B^+\to\pi^+\pi^0)|~.
\eeq
SU(3) breaking in this amplitude is often assumed to be given by the factor 
$f_K/f_\pi=1.193\pm 0.006$~\cite{Rosner:2008yu}. Here we introduce a parameter 
$\xi_{T+C}=1.0\pm 0.2$ which represents an uncertainty in this factor.  The
weak phase of $T+C$ is Arg($V^*_{ub} V_{us}) = \gamma$, where $\gamma = (65 \pm
10)^\circ$~\cite{CKM}. We take its strong phase to be zero by convention.
All other strong phases will be taken in the range $(-\pi,\pi)$. 
The penguin amplitude $P$ dominating $B\to K\pi$ decays carries the weak phase 
Arg$(V^*_{tb} V_{ts}) =\pi$. Its strong phase relative to that of $T+C$ will be denoted 
$-\delta_c$~\cite{Gronau:2001cj}. Thus 
\beq
T+C=|T+C|e^{i\gamma}~,~~~~~~~~~~~P=-|P|e^{-i\delta_c}~.
\eeq

The electroweak penguin contribution $P_{\rm EW}^C + P_{\rm EW}$ was shown in 
Refs.\ \cite{Neubert:1998pt} and \cite{Gronau:1998fn} to have the same strong phase 
as $T+C$ in the SU(3) symmetry limit.  In this limit the ratio of these two amplitudes is 
given numerically in terms of ratios of CKM factors and Wilson coefficients,
$(P_{\rm EW}+P_{\rm EW}^C)/(T+C)= -0.66\xi_{EW}e^{-i\gamma}$.
The parameter $\xi_{EW}$ includes an uncertainty from SU(3) breaking, which we will
take as $\xi_{EW}=1.0\pm 0.2$, and a smaller uncertainty from CKM factors. We neglect 
a potential small strong phase of $\xi_{EW}$ which has a negligible effect on our analysis 
below. Thus we have an amplitude triangle relation,
\beq \label{eqn:triangle}
A_{00} + A_{+-} = A_{3/2} = -|T+ C|\left(e^{i \gamma} - 0.66\xi_{EW}\right)~,
\eeq
and a similar relation for the CP-conjugate amplitudes in which 
the sign of $\gamma$ is reversed. 

In order to visualize the geometric construction of the triangle
(\ref{eqn:triangle}) and its CP-conjugate, 
as proposed in Ref.~\cite{Fleischer:2008wb} but with realistic quantities
including the restricted range (\ref{pred}) for $A_{CP}(K^0\pi^0)$, 
we express all branching ratios in units of $10^{-6}$,
and take amplitudes as their square roots.  (We first divide $B^+$ branching
ratios by the lifetime ratio $\tau(B^+)/\tau(B^0) = 1.071 \pm 0.009$
\cite{HFAG} to compare them with $B^0$ branching ratios.)  The central values
of $|T+C|$ for $\xi_{T+C}=1$ and the squares $|A_{ij}|^2$ and
$|\bar A_{ij}|^2$,  based on central values of the rates and CP asymmetries in 
Table \ref{tab:data}, are
\bea
|T+ C| & = & 0.900~, \nonumber \\
|A_{00}|^2     & = & 2(9.8)(1 + 0.14) = 22.3~, \nonumber \\
|A_{+-}|^2     & = & (19.4)(1 + 0.097) = 21.3~, \nonumber \\
|\bar A_{00}|^2     & = & 2(9.8)(1 - 0.14) = 16.9~, \nonumber \\
|\bar A_{-+}|^2     & = & (19.4)(1 - 0.097) = 17.5~.
\label{eqn:amps}
\eea

\begin{figure}
\begin{center}
\includegraphics[width=0.9\textwidth]{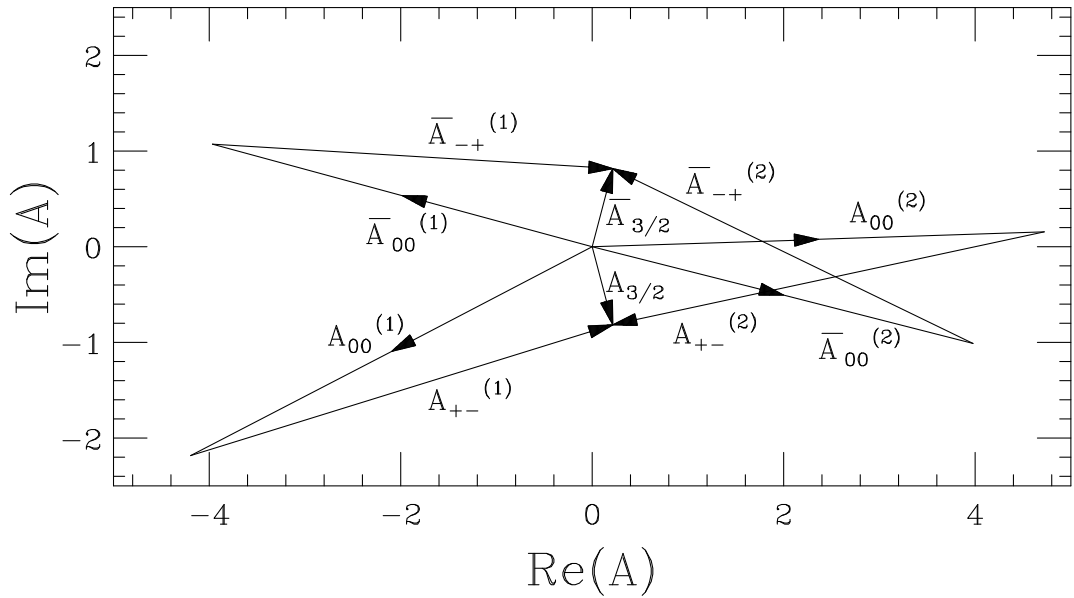}
\end{center}
\caption{Triangles relating amplitudes for $B^0 \to K^0 \pi^0$ and $B^0 \to
K^+ \pi^-$ to the amplitude $A_{3/2}$, and triangles for the corresponding
charge-conjugate processes.
\label{fig:tri}}
\end{figure}

Solutions for the amplitude triangle (\ref{eqn:triangle}) and its CP-conjugate
may be obtained analytically by solving simple quadratic equations for the
central values of the parameters which fix $A_{3/2}$ in (\ref{eqn:triangle}),
$\xi_{EW}=1$, $\gamma=65^\circ$.  The quadratic
equation for each triangle has two solutions, which can be visualized by
flipping the triangle around the side $A_{3/2}$ or $\bar A_{3/2}$ which is
kept fixed.  One thus obtains a total of $2 \times 2 = 4$ solutions, of which
two are illustrated in Fig.\ \ref{fig:tri}.  The other two solutions correspond
to flipping one triangle but not the other.

We have chosen to express the triangles with $A_{00}$ or $\bar A_{00}$
emanating from the origin, in order to illustrate the relative phase of
$A_{00}$ and $\bar A_{00}$ which will be important in the evaluation
of $\skp$.  This relative phase vanishes in the limit of pure penguin dominance
and is expected to be smaller than $\pi/2$ when including small 
color-suppressed tree and electroweak penguin contributions in $A_{00}$. 
This feature holds true for the two illustrated solutions but excludes
the two solutions with one triangle flipped, for which the relative phase between
$A_{00}$ and $\bar A_{00}$ is larger than $\pi/2$. 

The expected value of $\skp$ is related to the magnitudes and phases of
$A_{00}$ and $\bar A_{00}$ in the following manner:
\beq
\skp = \frac{2|A_{00} \bar A_{00}|}{|A_{00}|^2+|\bar A_{00}|^2}
\sin(2 \beta +  \phi_{00})~.
\label{eqn:skp}
\eeq
The correction $\phi_{00}\equiv {\rm Arg}(A_{00} \bar A_{00}^*)$ to $2 \beta$
is found to be positive for both of the displayed solutions. It is quite large,
$\phi_{00}=42.6^\circ$ corresponding to $\skp=0.99$, for the solution (1) with
negative real values of the amplitudes $A_{00}$ and $\bar A_{00}$ and smaller,
$\phi_{00}=16.1^\circ$ corresponding to  $\skp=0.85$, for the solution (2) with
positive real values.  Since $A_{00}$ is dominated by the penguin amplitude,
$P=-|P|\exp(-i\delta_c)$, solution (1) corresponds to $\cos\delta_c >0$
($|\delta_c|<\pi/2$) while solution (2) involves $\cos\delta_c<0$ ($|\delta_c|>
\pi/2$).

In order to exclude solution (2) one would have to show unambiguously that 
$\cos\delta_c>0$ or $|\delta_c|<\pi/2$, where $\delta_c$ is the strong phase 
difference between $T+C$ and $P$.  A most direct proof for $\cos\delta_c>0$
would need an observation of destructive interference between $P$ and $T+C$
in the CP-averaged decay rate of $B^+\to K^+\pi^0$ normalized by that of
$B^+\to K^0\pi^+$. However, this interference is cancelled by constructive 
interference of $P$ and $P_{EW}+P^C_{EW}$~\cite{Gronau:2006ha}.  Arguments for
small strong phase differences including $\delta_c$ have been presented in
studies of $B\to K\pi$ and $B\to\pi\pi$ based on  a heavy quark expansion
\cite{Beneke:1999br}.  These arguments failed, however, when
predicting a very small phase Arg$(C/T)$.  This would imply  $A_{CP}(K^+\pi^0)
<A_{CP}(K^+\pi^-)$, contrary to the two asymmetries quoted in Table
\ref{tab:data}, which show that this phase is not very 
small and must be negative (see argument below~\cite{Gronau:2006ha}.)
A small value of $\delta_c$ ($|\delta_c|<30^\circ$) was obtained in global
flavor SU(3) fits to decay rates and CP asymmetries measured for $B\to K\pi$
and $B\to\pi\pi$~\cite{Chiang:2004nm,Chiang:2006ih}. Within these fits it is
difficult to pinpoint a small subset of $B\to K\pi$ measurements forcing a
small value for $\delta_c$. The purpose of the subsequent discussion is to
prove $\cos\delta_c>0$ using a series of arguments based on specific
measurements, stressing the minimal use of untested assumptions about 
flavor SU(3).

A strong phase which is more directly accessible to experiment than $\delta_c$
is $\delta$, the strong phase
of $T$ relative to that of $P$.  This phase occurs in the amplitude for $B^0\to
K^+\pi^-$.  Its cosine term appears in the ratio $R$ of CP-averaged decay rates
for this process and $B^+\to K^0\pi^+$~\cite{Fleischer:1997um,Gronau:1997an}.
Neglecting $P^C_{EW}$ and $A$ terms in these amplitudes, one would expect $R$
to be smaller than one for $\cos\delta>0$ and larger than one for $\cos\delta
<0$. The current value $R=0.899\pm 0.048$, obtained from branching ratios in
Table \ref{tab:data} and the above-mentioned ratio of $B^+$ and $B^0$
lifetimes, favors $\cos\delta>0$ over $\cos\delta<0$. This evidence is
statistically limited and may suffer from $P^C_{EW}$ corrections in $B^0\to
K^+\pi^-$.  The negative asymmetry $A_{CP}(K^+\pi^-)=-0.097 \pm 0.012$ proves
unambiguously that $\delta$ is positive.
 
An argument proving $|\delta|<\pi/2$ unambiguously is based on the
time-dependent CP asymmetry parameter $\spp$ in $B^0 \to \pi^+ \pi^-$.
Assuming flavor SU(3), the ratio of penguin and tree amplitudes and their
relative phase are equal in this process to those in $B^0\to K^+\pi^-$, up to
CKM factors defining the ratios of amplitudes.  Neglecting small $W$-exchange
and penguin annihilation contributions (the resulting systematic uncertainty
introduced by this approximation is taken as part of an uncertainty due to
SU(3) breaking
mentioned below), one has~\cite{Gronau:2002qj}
\beq\label{Spp} 
\spp  =  \frac{\sin 2\alpha + 2r\cos\delta\sin(\beta-\alpha) -  
r^2\sin 2\beta}{1 - 2r\cos\delta\cos(\beta + \alpha) + r^2}~, 
\eeq
where $\alpha=\pi-\beta-\gamma$ and $r$ is the ratio of penguin and tree
amplitudes in $B^0\to\pi^+\pi^-$.
In the absence of a penguin amplitude one has $\spp=\sin 2\alpha$, and to first 
order in the ratio $r$ one finds
\cite{Gronau:2007af}
\beq
\spp = \sin 2 \alpha + 2 r \cos \delta \sin(\beta + \alpha) \cos 2 \alpha~.
\eeq
BaBar~\cite{Aubert:2007mj} and Belle~\cite{Ishino:2006if} find the same value
for this quantity; the average is large and negative\cite{HFAG}, $\spp = -0.61
\pm 0.08$.  Since $\alpha=\pi-\beta-\gamma\simeq \pi/2$~\cite{CKM} one has
$\sin 2 \alpha \simeq 0$ and $\cos 2 \alpha \simeq -1$, 
while $\sin(\beta + \alpha) > 0$, which implies $\cos \delta > 0$.  

A detailed analysis using the exact expression (\ref{Spp}) and measurements of
$\spp$ and a second asymmetry $\cpp\equiv -A_{CP}(\pi^+\pi^-)$ confirmed this
conclusion obtaining a value $\delta=(33\pm 7^{+8}_{-10})^{\circ}$
\cite{Gronau:2007af}.  The first error is experimental, while the second is
associated with a systematic uncertainty in flavor-SU(3) breaking. The positive
sign of $\delta$, following from the negative averaged $\cpp$, agrees with the
negative value of the measured $A_{CP}(K^+\pi^-)$. 
The two CP rate asymmetries are equal within experimental errors and 
have opposite signs~\cite{Deshpande:1994ii,Gronau:1995qd}. Expressed in 
units of $10^{-6}$ they are  $\Delta(K^+\pi^-)=-1.88\pm 0.24=-\Delta(\pi^+\pi^-)=
-1.96\pm 0.37$~\cite{HFAG}. This confirms the flavor SU(3) assumption for equal 
ratios of penguin and tree amplitudes and equal relative strong phases 
in these two processes. A difference of $180^\circ$ 
between the two phases, which would not affect the equality of
CP rate asymmetries, is extremely unlikely. The property $|\delta|<
\pi/2$ implies constructive 
(destructive) interference between $T$ and $P$ in the CP averaged
rate for $B^0\to\pi^+\pi^-$ ($B^0\to K^+\pi^-$).

\begin{figure}
\begin{center}
\includegraphics[width=0.8\textwidth]{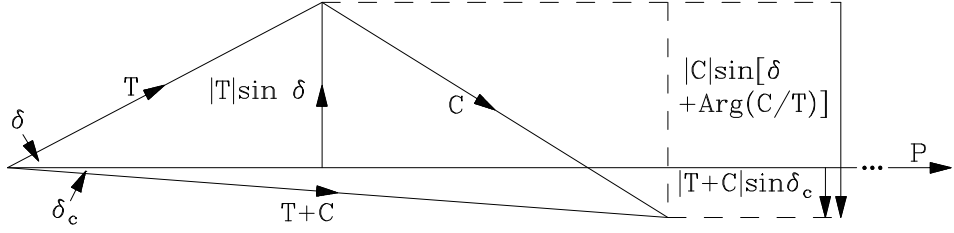}
\end{center}
\caption{Illustration of relative strong phases of $T$, $C$, and $P$ in
$B \to K \pi$ decays and the construction leading to Eq.\ (\ref{eqn:tri}).
Here $\delta = {\rm Arg}(T/P)$; $\delta_c = {\rm Arg}[(T+C)/P]$.
\label{fig:tctri}}
\end{figure}

In order to constrain $\delta_c$ (the strong phase difference between $T+C$
and $P$), using the above range for $\delta$ (the strong phase difference
between $T$ and $P$), one needs information about the strong phase of the ratio
$C/T$. The observation $A_{CP}(K^+\pi^0)>A_{CP}(K^+\pi^-)$ implies that
Arg$(C/T)$ is negative and larger in magnitude than $\delta$
\cite{Gronau:2006ha}. A simple proof of this behavior, for terms in the two
asymmetries which are linear in $|T+C|/|P|$ and $|T|/|P|$, respectively,
follows from the geometrical identity
\beq \label{eqn:tri}
 |T+C| \sin \delta_c = |T| \sin \delta + |C| \sin [\delta + {\rm Arg}(C/T) ]
\eeq
illustrated in Fig.\ \ref{fig:tctri}.
The amplitudes $T + C$ interfere constructively in $B^+\to \pi^+\pi^0$.  This
follows from the observation that $2\b(B^+\to\pi^+\pi^0)>\b(B^0\to\pi^+\pi^-)$
\cite{HFAG}, and the above-mentioned constructive interference of $T$ and $P$
in $B^0\to\pi^+\pi^-$.  Thus $-\pi/2 < {\rm Arg}(C/T) < -\delta<0$ which
implies geometrically $-\pi/2 < \delta_c < \delta< \pi/2$, without making any
assumption about the magnitude $|C/T|$. This concludes the proof of $\cos\delta_c>
0$ which excludes solution (2) in Fig.~1.

It is the large value of $\phi_{00}\equiv {\rm Arg}(A_{00} \bar A_{00}^*)$ in 
solution (1) in Fig.~1 which is thus responsible
for boosting the expected value of $\skp$ from its penguin-dominated value
of $\sin 2 \beta \simeq 0.68$ to a value very close to 1.  We now explore the
sensitivity of this effect to small changes in experimental inputs.

\begin{figure}
\mbox{\includegraphics[width=0.370\textwidth]{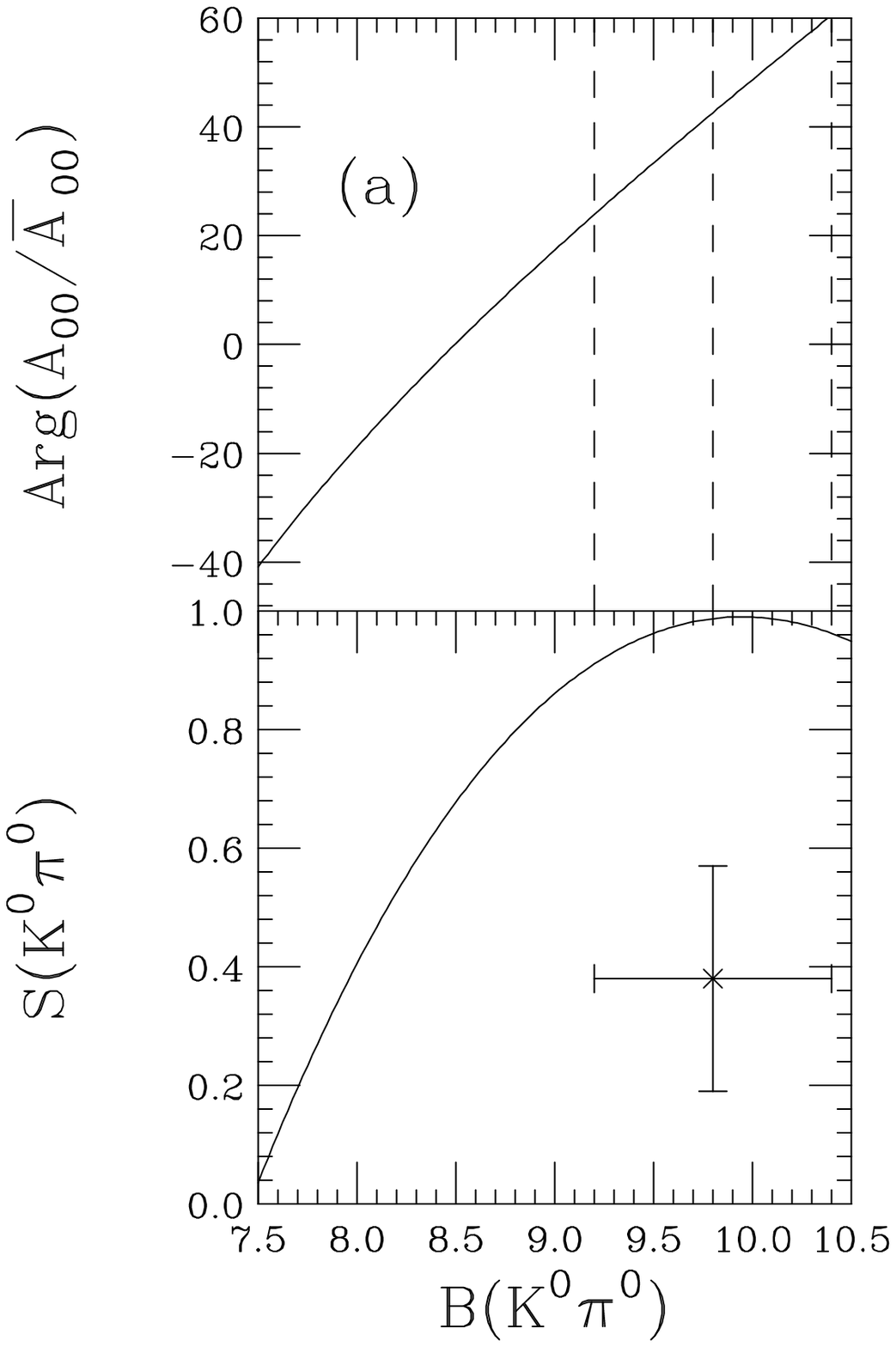}
      \includegraphics[width=0.300\textwidth]{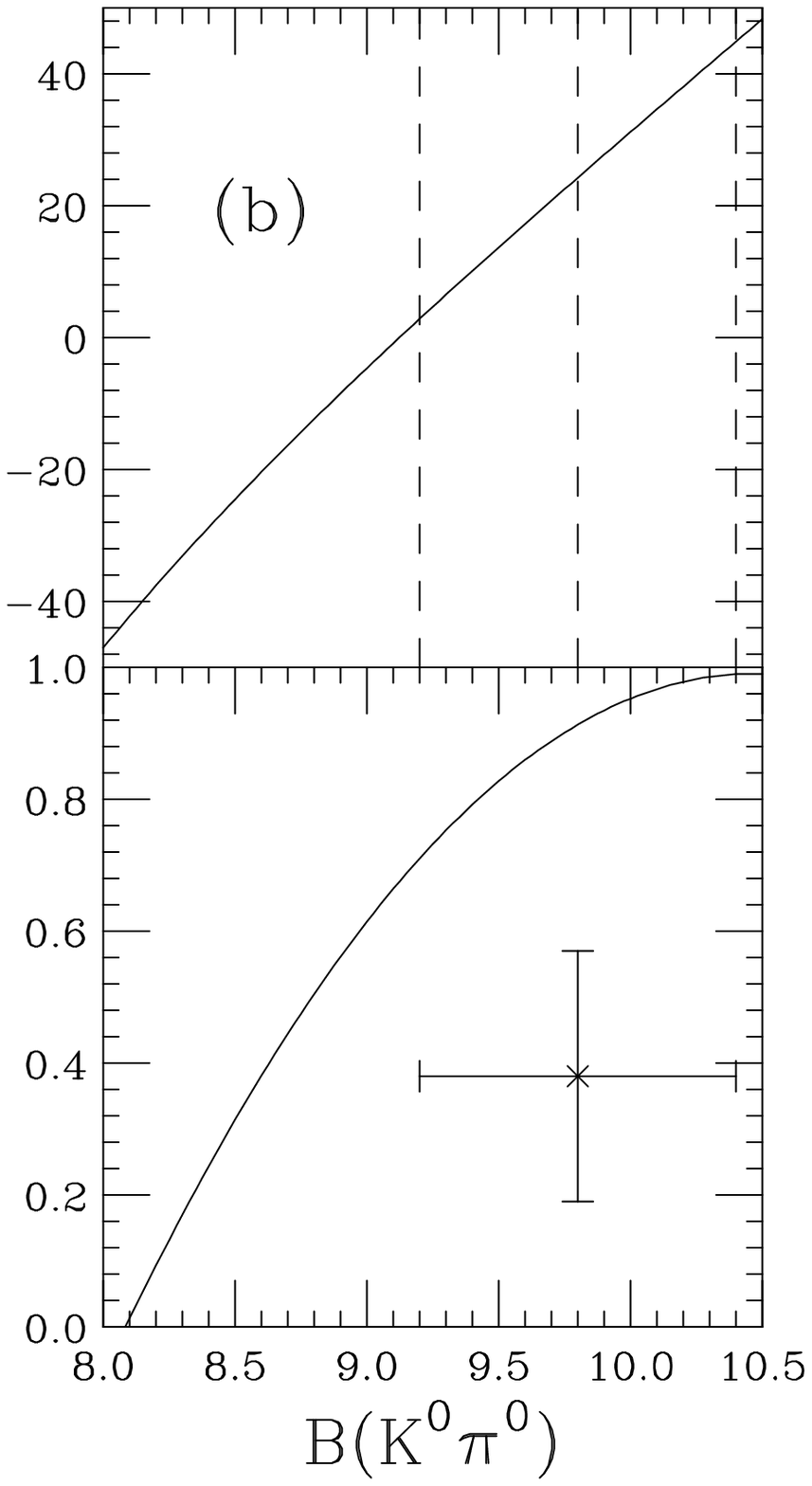}
      \includegraphics[width=0.300\textwidth]{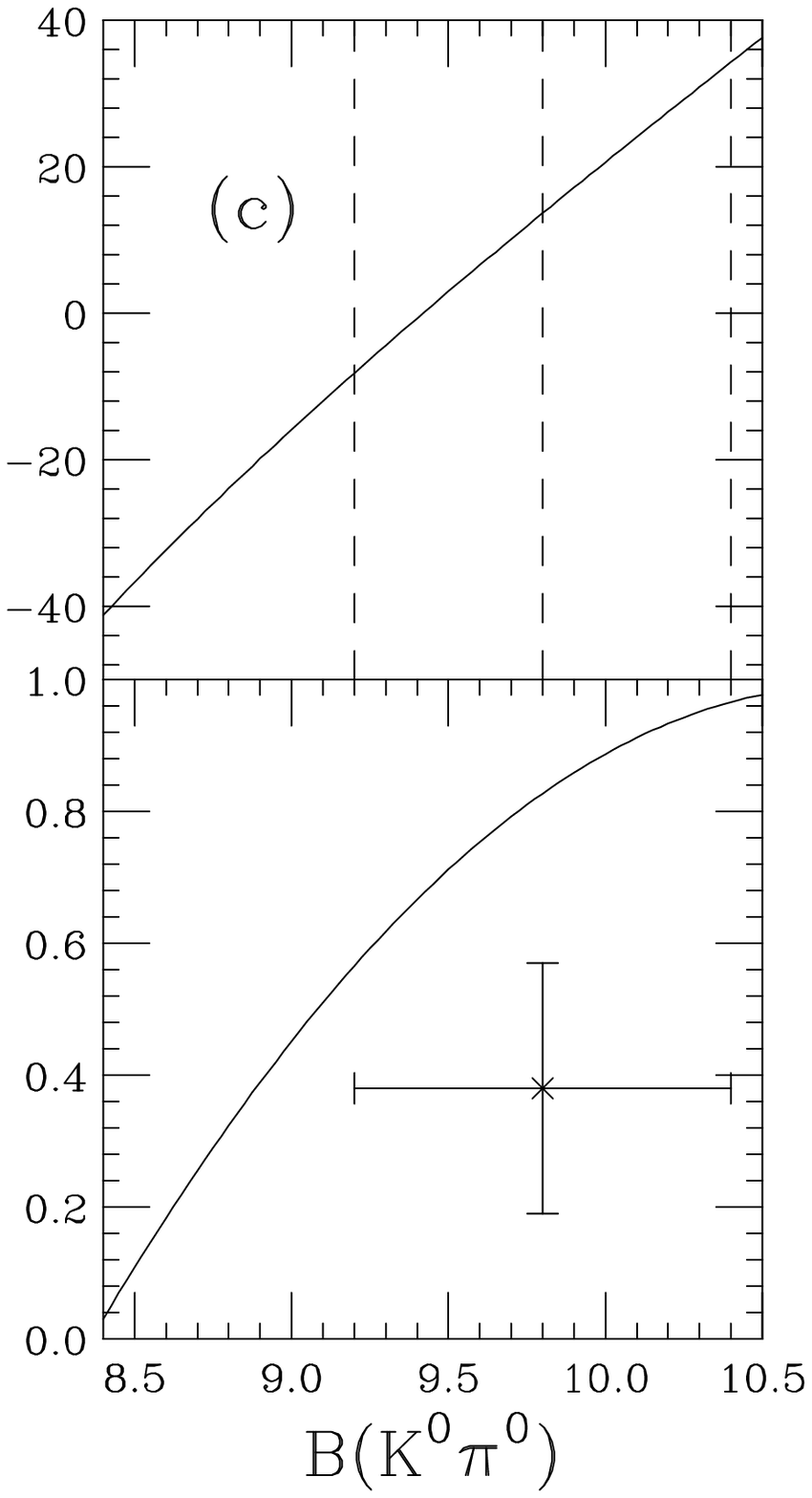}}
\caption{Dependence of Arg($A_{00}/\bar A_{00})$ and $\skp$ on B$(K^0 \pi^0)
\equiv \b(B^0 \to K^0 \pi^0)$.  Vertical dashed lines in top panel show central
value and $\pm 1\sigma$ errors of B$(K^0\pi^0)$.  The plotted point on the
lower panels shows the experimental values.  (a) All
parameters as in text; (b) same as (a), but $\gamma = 55^\circ$;
(c) same as (b), but $\b(B^0 \to K^+ \pi^-) = 20 \times 10^{-6}$.
\label{fig:var}}
\end{figure}
\begin{table}
\caption{Comparison of sensitivity of $\phi_{00} \equiv {\rm Arg}(A_{00}
\bar A_{00}^*)$ (in degrees) and $\skp$ to various parameters.
\label{tab:comp}}
\begin{center}
\begin{tabular}{c c c c c} \hline \hline
Parameter & \multicolumn{2}{c}{$-1\sigma$} & \multicolumn{2}{c}{$+1\sigma$} \\
                        & $\phi_{00}$ & $\skp$ & $\phi_{00}$ & $\skp$ \\ \hline
$\b(B^0 \to K^0 \pi^0)$ &  23.9  &  0.911 &  60.6  &  0.963 \\
$\gamma$                &  24.3  &  0.913 &  59.4  &  0.967 \\ 
$\b(B^0 \to K^+ \pi^-)$ &  52.0  &  0.986 &  33.3  &  0.962 \\
$\xi_{T+C}$             &  41.0  &  0.985 &  44.4  &  0.989 \\
$\xi_{EW}$              &  26.3  &  0.926 &  58.0  &  0.972 \\ \hline \hline
\end{tabular}
\end{center}
\end{table}

We find the greatest sensitivity of $\skp$ is to variations of the branching
ratio B$(K^0 \pi^0)$ $ \equiv  \b(B^0 \to K^0 \pi^0)$.  In Fig.\
\ref{fig:var}(a) we plot $\phi_{00}$ and $\skp$ versus B$(K^0\pi^0)$ for
nominal values of the parameters noted in the text.  We note that $\skp$ drops
from a value of 0.99 at the central value of B$(K^0\pi^0)$
to 0.91 and 0.72 at $-1\sigma$ and $-2\sigma$ below the central value.
We next vary $\gamma$ within its $1 \sigma$ limits to $55^\circ$ [Fig.\
\ref{fig:var}(b)].  The experimental values become considerably more compatible
with the Standard Model predictions, and even more so if $\b(B^0 \to K^+
\pi^-)$ is increased by $1 \sigma$ to $20 \times 10^{-6}$ [Fig.\
\ref{fig:var}(c)].  In Figs.\ 3 the quantity $\phi_{00}$ is more
sensitive than $\skp$ to variations in $\b(B^0 \to K^0 \pi^0)$, $\gamma$, and
$\b(B^0 \to K^+ \pi^-)$.  For the central value of $\phi_{00}$, $\skp$ is
very close to its maximum value, so it is only for considerably lower values
of $\phi_{00}$ that $\skp$ becomes sensitive to these parameters.  

In Table \ref{tab:comp} we summarize the effects on $\phi_{00}$ and $\skp$ of
varying $\b(B^0 \to K^0 \pi^0)$, $\gamma$, and $\b(B^0 \to K^+ \pi^-)$ by $\pm
1 \sigma$ around their central values.  
(See Table \ref{tab:data}; we are taking $\gamma = (65 \pm
10)^\circ$.)  A possible effect combining these three errors is seen in Fig.\
\ref{fig:var}(c).  We also include the effects of $\pm 1 \sigma$ variations of
$\xi_{T+C} = 1.0 \pm 0.2$ and $\xi_{EW} = 1.0 \pm 0.2$.  For nominal values of
the parameters, one has $\phi_{00}= 42.6^\circ$ and $\skp =
0.987$.  Table \ref{tab:comp} indicates the greatest sensitivity of $\phi_{00}$
to $\b(B^0 \to K^0 \pi^0)$, followed by $\gamma$ and $\xi_{EW}$.  There is
relatively little sensitivity to $\xi_{T+C}$.

Other variations are found to have a negligible effect on $\skp$.  
This includes the asymmetry $A_{CP}(B^0\to K^+\pi^-)$, which involves 
a very small experimental error, and  $A_{CP}(B^0\to K^0\pi^0)\equiv -\ckp$,
which is predicted in (\ref{pred}) with a small uncertainty.  A large variation
in this asymmetry would in any case have little effect on $\skp$, as a geometric
construction similar to that in Fig.\ \ref{fig:tri} illustrates.  The
phases of $A_{00}$ and $\bar A_{00}$ are found to shift nearly together,
so that the correction to $\sin 2 \beta$ in Eq.\ (\ref{eqn:skp}) changes
very little.  This insensitivity to $\ckp$ is displayed for the favored
$\skp$ solution in Ref.\ \cite{Fleischer:2008wb}, 
where $\ckp$ is left unconstrained disregarding the sum rule (\ref{SRDelta}).

Thus the  possibility that the above calculation of 
$\skp$ in the Standard Model differs both from
its penguin-dominated value of $\sin 2 \beta \simeq 0.68$ and from the data
remains intriguing.  However, for it to become a robust conclusion
about the presence of new physics, accuracies of measurements 
of the $B^0$ branching ratios  to $K^0 \pi^0$ and $K^+ \pi^-$ and of the CKM angle 
$\gamma$ need to be improved.  We look forward to such advances in future data,
and to more precise measurements of the two asymmetries $\ckp$ and $\skp$
in $B^0\to K^0\pi^0$. 

M.G. would like to thank the Enrico Fermi Institute at 
the University of Chicago for its kind and generous hospitality. We thank 
Masashi Hazumi for useful discussions, and Dan Pirjol and Jure Zupan for 
helpful communications.  This work was supported in part by 
the United States Department of Energy through Grant No.\ DE FG02 90ER40560.

\medskip
{\bf Note added:}
The measurements of $\ckp$ and $\skp$ given in 
Table  \ref{tab:ckpskp} have been updated very recently  by the BaBar and Belle collaborations.
New results and their averages are summarized in Table \ref{tab:ckpskp-update}.
%
\begin{table}[h]
\caption{Updated measurements of $\ckp$ and $\skp$.
\label{tab:ckpskp-update}}
\begin{center}
\begin{tabular}{c c c} \hline \hline
Ref.\ & $\ckp$ & $\skp$ \\ \hline
BaBar \cite{Hirschauer} & $0.13 \pm 0.13 \pm 0.03$ & $0.55 \pm 0.20 \pm 0.03$ \\
Belle \cite{Dalseno} & $-0.14 \pm 0.13 \pm 0.06$ & $0.67 \pm 0.31 \pm 0.08$ \\
Average & $0.00 \pm 0.10$ & $0.58 \pm 0.17$ \\ \hline
\end{tabular}
\end{center}
\end{table}
The averaged value of $\ckp$ agrees with the prediction (\ref{pred}) within $1.4\sigma$,
while $\skp$ is now consistent with $\sin 2\beta$ and somewhat larger values. 
Recent updates by BaBar of the branching ratio for $B^0\to K^0\pi^0$ and the CP 
asymmetry in $B^0\to K^+\pi^-$~\cite{Aubert:2008sb} do not affect significantly the 
corresponding two averaged values in Table \ref{tab:data}.

\end{document}